\documentclass[traditabstract]{aa}  
\usepackage{natbib}
\usepackage{graphicx}
\usepackage{amsmath}
\usepackage{txfonts}
\usepackage{amssymb}
\begin{document} 

   \title{Compact steep-spectrum sources as the\\
  parent population of flat-spectrum radio-loud NLS1s}


   \author{M. Berton
	\inst{1}\thanks{marco.berton.1@studenti.unipd.it}
 	\and A. Caccianiga\inst{2}
	\and L. Foschini\inst{2}
	\and B.~M. Peterson\inst{3}
	\and S. Mathur \inst{3}
	\and G. Terreran\inst{4,5} \and
	\\
	S. Ciroi\inst{1} 
	\and E. Congiu\inst{1}
	\and V. Cracco\inst{1} 
	\and M. Frezzato\inst{1}
	\and G. La Mura\inst{1}
	\and P. Rafanelli\inst{1}	
         }

   \institute{$^{1}$ Dipartimento di Fisica e Astronomia "G. Galilei", Universit\`a di Padova, Vicolo dell'Osservatorio 3, 35122 Padova, Italy;\\
 $^{2}$ INAF - Osservatorio Astronomico di Brera, via E. Bianchi 46, 23807 Merate (LC), Italy;\\
	$^{3}$ Department of Astronomy and Center for Cosmology and AstroParticle Physics, The Ohio State University, 140 West 18th Avenue, Columbus, OH 43210, USA;\\
	$^{4}$ Astrophysics Research Centre, School of Mathematics and Physics, Queen's University Belfast, Belfast BT7 1NN, UK;\\
	$^{5}$ INAF - Osservatorio Astronomico di Padova, Vicolo dell'Osservatorio 5, 35122 Padova, Italy.\\
             }

\authorrunning{M. Berton et al.}
\titlerunning{CSS sources as the parent population of F-NLS1s}

\abstract{Narrow-line Seyfert 1 galaxies (NLS1s) are an interesting subclass of active galactic nuclei (AGN), which tipically does not exhibit any strong radio emission. Seven percent of them, though, are radio-loud and often show a flat radio-spectrum (F-NLS1s). This, along to the detection of $\gamma$-ray emission coming from them, is usually interpreted as a sign of a relativistic beamed jet oriented along the line of sight. An important aspect of these AGN that must be understood is the nature of their parent population, in other words how do they appear when observed under different angles. In the recent literature it has been proposed that a specific class of radio-galaxies, compact-steep sources (CSS) classified as high excitation radio galaxies (HERG), can represent the parent population of F-NLS1s. To test this hypothesis in a quantitative way,in this paper we analyzed the only two statistically complete samples of CSS/HERGs and F-NLS1s available in the literature. We derived the black hole mass and Eddington ratio distributions, and we built for the first time the radio luminosity function of F-NLS1s. Finally, we applied a relativistic beaming model to the luminosity function of CSS/HERGs, and compared the result with the observed function of F-NLS1s. We found that compact steep-spectrum sources are valid parent candidates and that F-NLS1s, when observed with a different inclination, might actually appear as CSS/HERGs.}

\keywords{Galaxies: Seyfert; galaxies: jets; galaxies: luminosity function; quasars: emission lines; quasars: supermassive black holes}
\maketitle

\newcommand{\kms}{km s$^{-1}$}
\newcommand{\ergs}{erg s$^{-1}$}
\renewcommand{\lll}{\mathcal{L}}
\section{Introduction}
Since their first designation as a subclass of active galactic nuclei (AGN) in 1985 \citep{Osterbrock85}, narrow-line Seyfert 1 galaxies (NLS1s) represented a source of new insights. By definition these AGN have a relatively low full width at half maximum (FWHM) of the permitted lines, specifically a FWHM(H$\beta$) $<$ 2000 \kms\ and a flux ratio of [O III]/H$\beta <$ 3 \citep{Osterbrock87, Goodrich89}. Moreover, the presence in the optical spectrum of strong Fe II multiplets shows that the broad-line region (BLR) of these AGN is directly visible. The narrowness of the permitted lines cannot therefore be interpreted as due to obscuration, but instead to a low rotational velocity around a relatively low mass central black hole (10$^{6-8}$ M$_\odot$, \citealp{Mathur00}). This low black hole mass, along with the high Eddington ratio \citep{Boroson92}, is sometimes interpreted as a consequence of the young age of these sources \citep{Grupe00, Mathur00}. Therefore, NLS1s may be rapidly growing AGN and, if so, they are an excellent proxy to test the evolution of galaxies at low redshifts. \par
Although they are typically radio-quiet, a fraction of NLS1s are actually radio loud (7\%) or even very radio loud (2.5\%, \citealp{Komossa06}). Few of these exhibit some extreme properties similar to those of blazars, such as flat radio spectra ($\alpha_\nu \leq 0.5$, with F$_\nu \propto \nu^{-\alpha_\nu}$) and high brightness temperatures \citep{Yuan08}. After the discovery of $\gamma$-ray emission coming from these sources and detection by the \textit{Fermi Gamma-ray Space Telescope} \citep{Abdo09a,Abdo09b,Abdo09c, Foschini10}, NLS1s became the third class of $\gamma$-ray emitting AGN with a relativistic beamed jet, in addition to BL Lacs and flat spectrum radio quasars (FSRQs). \par
According to the unified model of radio-loud AGN, for each highly beamed source there are about 2$\Gamma^2$ misaligned sources ($\Gamma$ is the bulk Lorentz factor of the jet), also known as the parent population. In particular, the parent populations of BL Lacs and FSRQs are thought to be FRI and FRII radio galaxies, respectively \citep{Urry95}, even though some exceptions are known \citep{Kollgaard92, Antonucci02}. In a more recent and less biased picture, the reviewed association with parent sources is between BL Lacs and low excitation radio-galaxies (LERG), and between FSRQs and high excitation radio galaxies (HERG; \citealp{Giommi12}). An akin picture for flat-spectrum radio-loud NLS1s (F-NLS1s) is not yet well established. \par
An itial hint regarding the parent population was provided by \citet{Foschini11, Foschini12}, who first proposed steep-spectrum radio-loud NLS1s (S-NLS1s) as parent sources. This hypothesis was further supported by \citet{Berton15a}, although a possible problem of numerical consistency was also pointed out, since the known S-NLS1s seem to be too few to represent the whole parent population. It is however possible that there is no need to include anything else in the parent population beside S-NLS1s. A GHz selected sample might show a lack of misaligned sources because the relativistic beaming increases the luminosity, and hence the visibility, of beamed sources. An example for this is shown by \citet{Urry95}. In their study on the radio luminosity functions (LFs), the observed densities of FSRQs and steep-spectrum radio-quasars (SSRQs) at 2.7 GHz are roughly the same, so one could expect the number of observed beamed and misaligned sources to be the same for NLS1s as well. \par
Alternatively the lack of sources might be real and could be explained with two different hypotheses \citep{Foschini10,Foschini11}. The first hypothesis is based on the fact that the radio morphology of F-NLS1s is extremely compact on a parsec scale \citep[e.g.,][]{Doi11}. Young sources, such as NLS1s, might have not developed radio lobes yet. The radio emission might be then strongly collimated and, when observed at large angles, the source would be invisible for present day observatories, appearing as a radio-quiet NLS1s (RQNLS1s). Another hypothesis is instead based on a different assumption on the nature of NLS1s. Some authors suggest that the narrowness of the permitted lines might be only an apparent effect \citep{Decarli08, Shen14}. If the BLR has a disk-like shape, when observed pole-on the permitted lines would show little rotational Doppler broadening, and would appear narrower than at higher inclination. The NLS1s would then be nearly pole-on AGN. Increasin their observing angle instead, the permitted lines would become broader, and the source might appear as a broad-line or a narrow-line radio-galaxy (BLRG/NLRG), whether the line of sight intercepts the torus or not. Typically, NLS1s are hosted in disk-like galaxies \citep{Crenshaw03}. If the host galaxy is also the same for F-NLS1s, the parent source would be disk-hosted BL/NLRG. These hypotheses were also investigated by \citet{Berton15a}, who suggested that disk RGs with high Eddington ratio and low black hole mass may be parent sources, while RQNLS1s are not very good candidates. \par
There is, in any case, another possibility that must be considered. If radio lobes are lacking, RQNLS1s belong to the parent population, but lobes might instead only be developed on small scale. In this case, the source could appear as a compact steep spectrum object (CSS). The CSS represent an important portion of radio sources, showing a radio spectrum peaked at $\sim$100 MHz and radio-jets entirely within the host galaxy \citep[see the review by][]{Odea98}. Often they are thought to be closely connected with Gigahertz peaked-spectrum sources (GPS), and a widely used theory to explain their nature is the youth scenario \citep{Fanti95}. Their age was determined in several ways, and found to be less than 10$^5$ years \citep{Owsianik98, Murgia99}. Their jets already developed radio lobes, and are typically still crossing the interstellar medium and interacting with it \citep[e.g.,][and references therein]{Morganti15}. As radio galaxies, they can be classified as HERG or LERG according, for example, to the ratio [O III]/H$\alpha$ \citep{Laing94}. The main difference between these classes is likely the accretion mechanism onto the black hole, with HERGs showing typically a more efficient accretion process \citep{Hardcastle07}. \par
A link between NLS1s and CSS was suggested by many authors \citep{Oshlack01, Komossa06, Gallo06a, Yuan08, Caccianiga14, Gu15, Schulz15}. All the characteristics of NLS1s indeed closely recall those of CSS. As previously mentioned they are often considered young sources and at the same time their jet, when present, appears to interact with the medium \citep{Marziani03, Komossa06}. RLNLS1s can be classified as HERG, having an efficient accretion mechanism and strong high-ionization lines. Therefore CSS/HERGs might be part of their parent population under the assumption that radio lobes are already developed in RLNLS1s.\par 
The aim of the present work is to study whether CSS/HERGs are suitable to be the parent population of F-NLS1s or not. The paper is organized as follows: in Sect.~2 we present the samples selection; in Sect.~3 we describe the black hole mass and Eddington ratio analysis; in Sect.~4 we present the V/V$_{max}$ test for the samples; in Sect.~5 we build the luminosity functions and study the incidence of relativistic beaming; in Sect.~6 we discuss our results; and, finally, in Sect.~7 we briefly summarize our work. Throughout this work we adopt a standard $\Lambda$CDM cosmology with H$_0 = 70$ \kms\ Mpc$^{-1}$, $\Omega_M = 0.3$ and $\Omega_\Lambda = 0.7$ \citep{Komatsu11}. 
\section{Samples}
\subsection{NLS1s}
Complete samples are required to carry out a reliable comparison between F-NLS1s and the putative parent population. The NLS1s sample must also have measured spectral indices, to select only flat-spectrum sources without including any S-NLS1s. The largest sample in the literature that meets our requirements is that of \citet{Yuan08}. It includes 23 very radio-loud NLS1s, and 21 out of 23 spectral indices are known. Their sample was selected from SDSS DR5, looking only for those sources whose radio loudness, calculated using the 1.4 GHz flux density, is above 100 at $z < 0.8$. The sample includes 15 flat-spectrum sources and 2 with unknown spectral index. \par
The Yuan sample should be statistically complete, because it is drawn from the already complete sample of \citet{Zhou06}. In any case, we independently tested its completeness. We first notice that \citet{Foschini15}, whose sample was selected with an accurate search in the literature, found all the very radio-loud objects in DR5 already included in the Yuan sample. The cumulative distribution of sources as a function of redshift nevertheless shows a flattening close to the upper $z$ limit. This flattening in the distribution is likely caused by the lack of classification. The signal to noise ratio (S/N) of the optical spectra worsen with increasing distance, and even if very radio-loud sources typically have bright optical lines (such as [O III], e.g. \citealp{debruyn78}), a correct classification is very difficult. \par
We tried to avoid this problem considering that the quasars distribution in SDSS appears to be complete up at 94.6\% up to magnitude i$<$19.1 \citep{Richards02}. In the Yuan sample, all the flat-spectrum sources but one match this magnitude criterium when $z < 0.6$. We therefore decided to use this threshold as the upper redshift limit for our sample. This allows us to have a good degree of completeness in our sample. Using these criteria, 13 F-NLS1s remain. We also consider one more source with undetermined spectral index, which meets the redshift criterion, to test the stability of our results.

\subsection{High excitation radio galaxies}
Our aim was to find CSS sources classified as HERGs, so we searched again in the literature for a suitable sample. We decided to use that of \citet{Kunert10a}, who selected a sample of 44 low-luminosity compact objects with a radio luminosity at 1.4 GHz lower than 10$^{26}$ W Hz$^{-1}$ (in a cosmology with H$_0$ = 100 \kms\ Mpc$^{-1}$ and q$_0$ = 0.5). In addition to this criterion, these sources have a flux density 70 mJy $\leq S_{1.4 \; GHz} \leq$ 1 Jy and a radio spectral index $\alpha_\nu > 0.7$ between 1.4 and 4.85 GHz. Their radio-selected sample was later cross-matched with the SDSS DR7 spectroscopic archive, finding 29 sources at z $<$ 0.9 \citep{Kunert10b}. Ten of these sources were classified as LERG, 12 as HERG, and 7 remained unclassified because they had a S/N $\lesssim 3$. We tested the completeness of this sample as before. As in the F-NLS1s sample, the cumulative distribution finds a drop in the source counts above z $\sim$ 0.6. Below this threshold, only one source is above the SDSS completeness limit of 19.1 mag. Therefore we decided to use the same limits again, this time considering only ten sources. Since both samples have the same redshift limit and they both have a lower limit in flux, including only bright radio sources, the comparison between them should be relatively unbiased. 

\subsection{Control sample}
As a control sample for the luminosity function, we decided to use the sample of 50 FSRQs used by \citet{Padovani92}, which in turn are drawn from the work of \citet{Wall85}. The sources have a flux density above 2 Jy at 2.7 GHz, and Galactic latitude $|b| >$ 10$^\circ$. They also have a spectral index $\alpha_\nu \leq 0.5$ between 2.7 and 5 GHz, and they were not classified as BL Lacs by \citet{Stickel91}. \citet{Padovani92} added one more source to the \citet{Wall85} sample, because of its high optical polarization. 

\section{Black hole mass}
\begin{table}[t!]
\centering
\caption{HERGs parameters. }
\begin{tabular}{l c c c}
\hline\hline
SDSS Name & $\log$M$_{BH}$ & $\log$L$_{bol}$ & $\log$Edd \\
\hline
SDSS J002833.42+005510.9 & 8.94 & 44.1 & -2.96 \\
SDSS J075756.71+395936.0 & 7.13 & 43.77 & -1.52 \\
SDSS J084856.57+013647.8 & 7.05 & 44.41 & -0.74 \\
SDSS J092607.99+074526.6 & 7.28 & 44.93 & -0.47 \\
SDSS J094525.90+352103.5 & 7.23 & 44.48 & -0.85 \\
SDSS J114311.01+053516.1 & 8.84 & 45.08 & -2.0 \\
SDSS J115727.61+431806.3 & 7.68 & 44.67 & -1.1 \\
SDSS J140416.35+411748.7 & 7.96 & 43.88 & -2.22 \\
SDSS J140942.44+360415.8 & 8.24 & 43.82 & -2.52 \\
SDSS J164311.34+315618.4 & 7.44 & 45.39 & -0.17 \\

\hline
\end{tabular}
\label{masse}
\tablefoot{Columns: (1) object SDSS name; (2) logarithm of the black hole mass in $M_\odot$; (3) logarithm of the bolometric luminosity in \ergs; (4) logarithm of the Eddington ratio.}
\end{table}
An important step to understand the relation between CSS/HERGs and F-NLS1s is to compare their black hole masses and Eddington ratio. Both these values were recently calculated for all our F-NLS1s by \citet{Foschini15}, therefore we adopt their values for our study. For CSS/HERGs we obtained the optical spectra from SDSS DR12. All of these sources were of type 2 or intermediate type AGN, therefore we could not use permitted lines to derive the black hole mass because the BLR is obscured. We then followed the procedure described by \citet{Berton15a} for type 2 and intermediate sources, deriving the stellar velocity dispersion $\sigma_*$ from the width of the [O III] lines core component. Once the blue wing is removed, the core component of [O III] should indeed be less affected by the jets/ISM interaction and typically dominated by the gravitational potential of the bulge stars \citep{Greene05}. This method provided good approximations for black hole mass both in elliptical- and disk-hosted radio galaxies. To obtain the bolometric luminosity, we used Eq.~7 of \citet{Berton15a}, i.e.,
\begin{equation}
\log\left(\frac{L_{\mathrm{bol}}}{\textrm{erg s}^{-1}}\right) = (7.54\pm9.07) + (0.88\pm0.22)\log\left(\frac{L_{\mathrm{[OIII]}}}{\textrm{erg s}^{-1}}\right) \, .
\end{equation}
 \par
The results are shown in Tab.~\ref{masse}. The logarithmic mean mass value for HERGs is 7.78 with a standard deviation of 0.66, while for F-NLS1s is 7.68 with a standard deviation of 0.44. The median values are 7.84 and 7.73, respectively. It is evident that the two distributions are very similar. We compared them by means of the Kolmogorov-Smirnov test (K-S). The null hypothesis is that the two distributions are originated from the same population. The rejection threshold for the K-S throughout this work is a p-value lower than 0.05. In the case of HERGs and F-NLS1s, we found a p-value of 0.95, thereby not allowing us to reject the null hypothesis. To directly compare this result with those found for other parent candidates by \citet{Berton15a}, we also evaluated the product 
\begin{equation}
\mathcal{P} = D_n \sqrt{\frac{nm}{n+m}} \; ,
\end{equation}
where $D_n$ is the deviation between the cumulative distributions and $n$ and $m$ are the number of elements in each sample. This value is useful to test the distance between the samples. A low value indicates that the null hypothesis can be rejected in the two samples. In our case this value is equal to 0.52, and the CSS/HERGs sample appears to be the closest to F-NLS1s and even closer than S-NLS1s. \par
\begin{figure}[t!]
\centering
\includegraphics[width=\hsize]{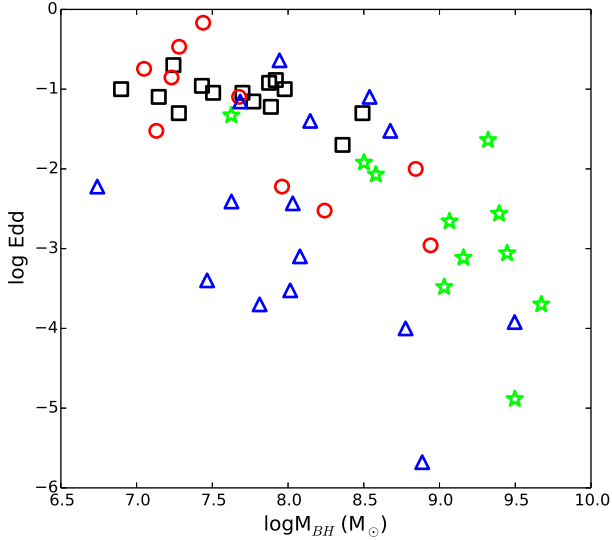} 
\caption{Logarithm of the BH mass vs. logarithm of the Eddington ratio. Black squares indicate F-NLS1s, red circles indicate CSS/HERGs, blue triangles indicate disk-hosted radio-galaxies and green stars indicate elliptical-hosted radio galaxies. The points of these last two samples are derived from \citet{Berton15a}.}
\label{massacc}
\end{figure}

The Eddington ratio distributions are also fairly close (K-S p-value 0.17, the null hypothesis is again the two distributions originated from the same population). The logarithmic median Eddington ratio is -1.31 for CSS/HERGs and -1.05 for F-NLS1s with a standard deviation of 0.89 and 0.23, respectively. These values are comparable to those of other NLS1s classes, but it is worth noting that the CSS/HERGs sample shows some outliers with lower accretion luminosities, hence a larger standard deviation. All these values are shown in Fig.~\ref{massacc} along with two more classes of sources already studied by \citet{Berton15a}, namely, disk-hosted and elliptical-hosted radio-galaxies. \par
The good overlap in mass between F-NLS1s and CSS/HERGs is visible and we also point out that disk RGs have a similar mass distribution to CSS/HERGs (K-S p-value 0.15). On the contrary, the black hole mass of elliptical radio-galaxies is much larger. The K-S confirms this difference providing a p-value of 3$\times$10$^{-3}$, which allows us to reject the null hypothesis of the two distributions originated from the same population. 
\section{V/V$_{max}$ test}
\label{VVmax}
Another step to understand the relation between these sources is to check whether evolution is present in our samples or not. An useful tool to test evolution is the so-called V/V$_{max}$ test \citep{Schmidt68}. By definition, V$_{max}$ is the volume within which a source of luminosity $L$ can be detected, while $V$ is the spherical volume associated with each source. The luminosity of a source of detected flux $F$ is $L = 4\pi d^2 F$, where $d$ is the luminosity distance. If the flux detection limit is F$_{min}$, the source can be detected up to 
\begin{equation}
d_{max} = \sqrt{\frac{L}{4\pi F_{min}}}
\end{equation}
which corresponds to a redshift $z_{max}$. For a nonevolving population, the ratio between the spherical volume V, corresponding to the object redshift and V$_{max}$, is expected to be uniformly distributed between 0 and 1 with an average value $\langle V/V_{max} \rangle = 0.5$. When $\langle V/V_{max} \rangle > 0.5$, the population is positively evolving with more (or more luminous) sources located at larger distances. If conversely $\langle V/V_{max} \rangle < 0.5$, the sample is negatively evolving. \par
To evaluate V$_{max}$ for each object, we used as $d_{max}$ the smaller value between those derived from the radio detection limit (1 mJy for F-NLS1s, 70 mJy for CSS/HERGs, and 2 Jy for FRSQs), the spectroscopic limit for quasars in SDSS DR7 (19.1 mag) and the redshift upper limit of each sample, z = 0.6. The CSS/HERGs sample has also an upper flux limit, which translates into a lower redshift limit z$_{min}$. Therefore in this case we used the modified version of the test over the accessible volume V$_a$ \citep{Avni80}, which is defined as
\begin{equation}
\frac{V}{V_a} = \frac{V - V_{min}}{V_{max} - V_{min}} \; ,
\end{equation}
where V$_{min}$ is the inaccessible inner part of the comoving volume due to z$_{min}$ and $V$ is the comoving volume of each source. The associated error in the V/V$_{max}$ test is $\sigma = 1/\sqrt{12N}$, where $N$ is the number of sources in each sample. To calculate both the luminosity distance from redshift and the comoving volume, we used the \texttt{Cosmolopy} tool developed for \texttt{Python}\footnote{http://roban.github.com/CosmoloPy/}. \par
The results are summarized in Tab.~\ref{vvmax}. The control sample of FSRQs shows a strong positive evolution at 5$\sigma$. This result is in agreement with that found by \citet{Padovani92} and many other authors. Indeed, FSRQs are known for having a strong evolution with time. Conversely, the V/V$_{max}$ result is consistent with the uniform distribution at 1$\sigma$ for both F-NLS1s and CSS/HERGs. In particular the result for F-NLS1s does not change whether the two sources with undetermined spectral index are included or not. This is an indication that, at least up to $z = 0.6$, these sources do not have a strong evolution. \par
We performed a K-S test between the observed V/V$_{max}$ distributions in our sample and the theoretical uniform distribution. The null hypothesis is that the observed distribution is drawn from a uniform distribution. As reported in Tab.~\ref{vvmax}, the test confirms all the previous results, showing that the only sample where the null hypothesis is rejected is the FSRQs sample. Therefore, while the luminosity function of FSRQs is corrected for evolution and reported to z = 0, those of F-NLS1s and CSS/HERGs are not.
\begin{figure}[t!]
\centering
\includegraphics[width=\hsize]{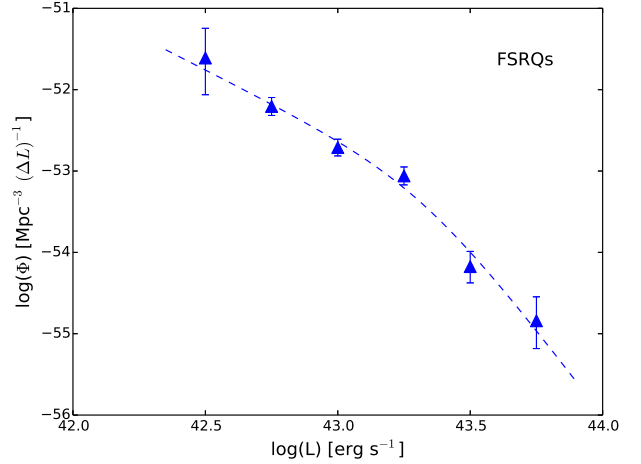} 
\caption{Monochromatic radio luminosity function of FSRQs control sample at 1.4 GHz. The dashed line is the best fit with a broken power law. }
\label{fsrq_lf}
\end{figure}
\begin{table}[t!]
\centering
\caption{Results of the V/V$_{max}$ test. The results for F-NLS1s are showed in two different ways: with or without 1 source with unknown spectral index. The former are indicated with an asterisk. }
\begin{tabular}{l c c c c c}
\hline\hline
Sample & N & V/V$_{max}$ & $\sigma$ & d & K-S \\
\hline
F-NLS1 & 13 & 0.55 & 0.08 & 0.63 & 0.52 \\
F-NLS1* & 14 & 0.58 & 0.08 & 1.00 & 0.26 \\
HERG & 10 & 0.54 & 0.09 & 0.44 & 0.72 \\
FSRQ & 50 & 0.70 & 0.04 & 5.00 & 2$\times$10$^{-9}$\\
\hline
\end{tabular}
\label{vvmax}
\tablefoot{Columns: (1) sample; (2) number of sources; (3) result of the test; (4) associated error with the test; (5) distance from uniform distribution in $\sigma$ units; (6) K-S test p-value against uniform distribution.}
\end{table}
\begin{figure}[!t]
\centering
\includegraphics[width=\columnwidth]{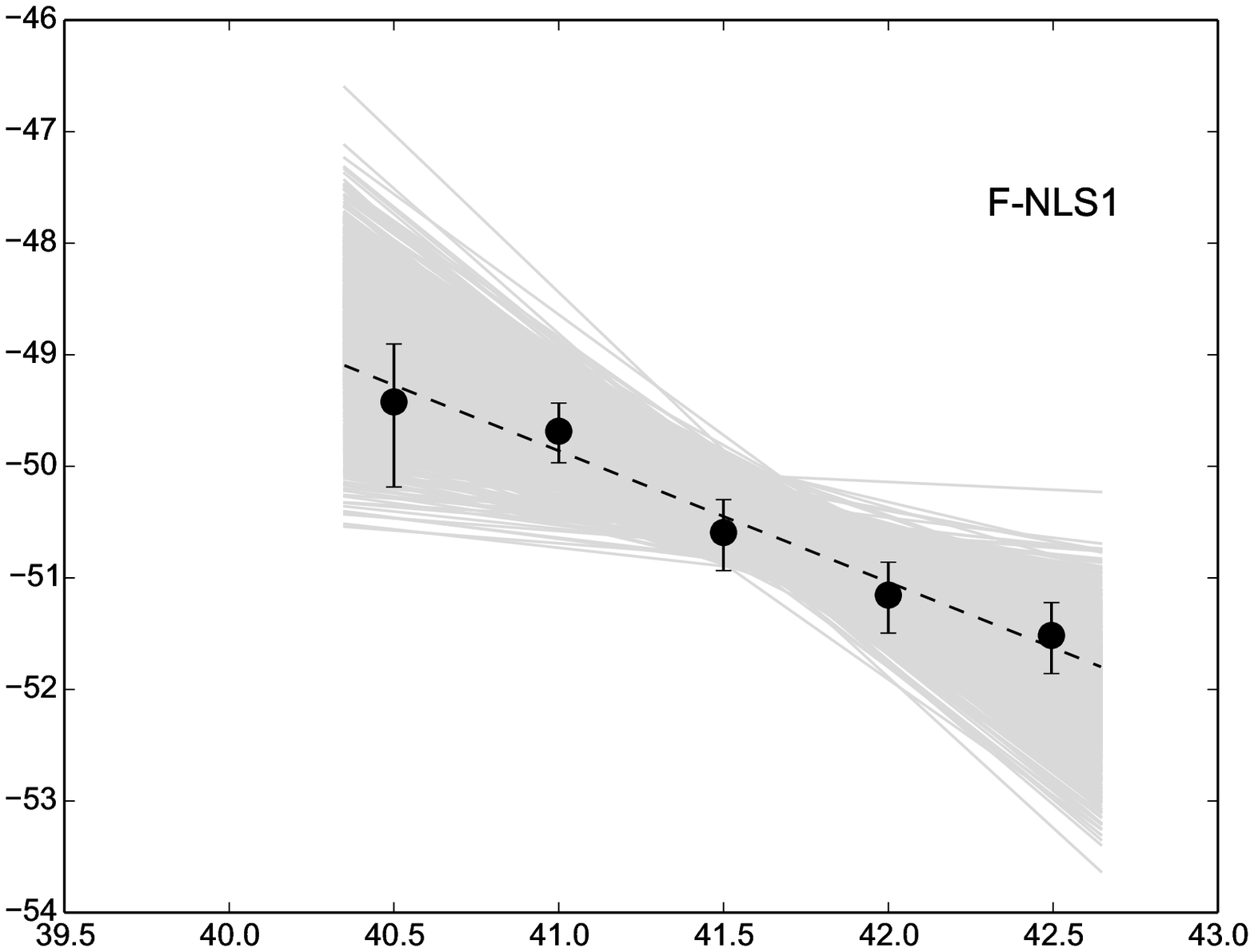} 
\includegraphics[width=\columnwidth]{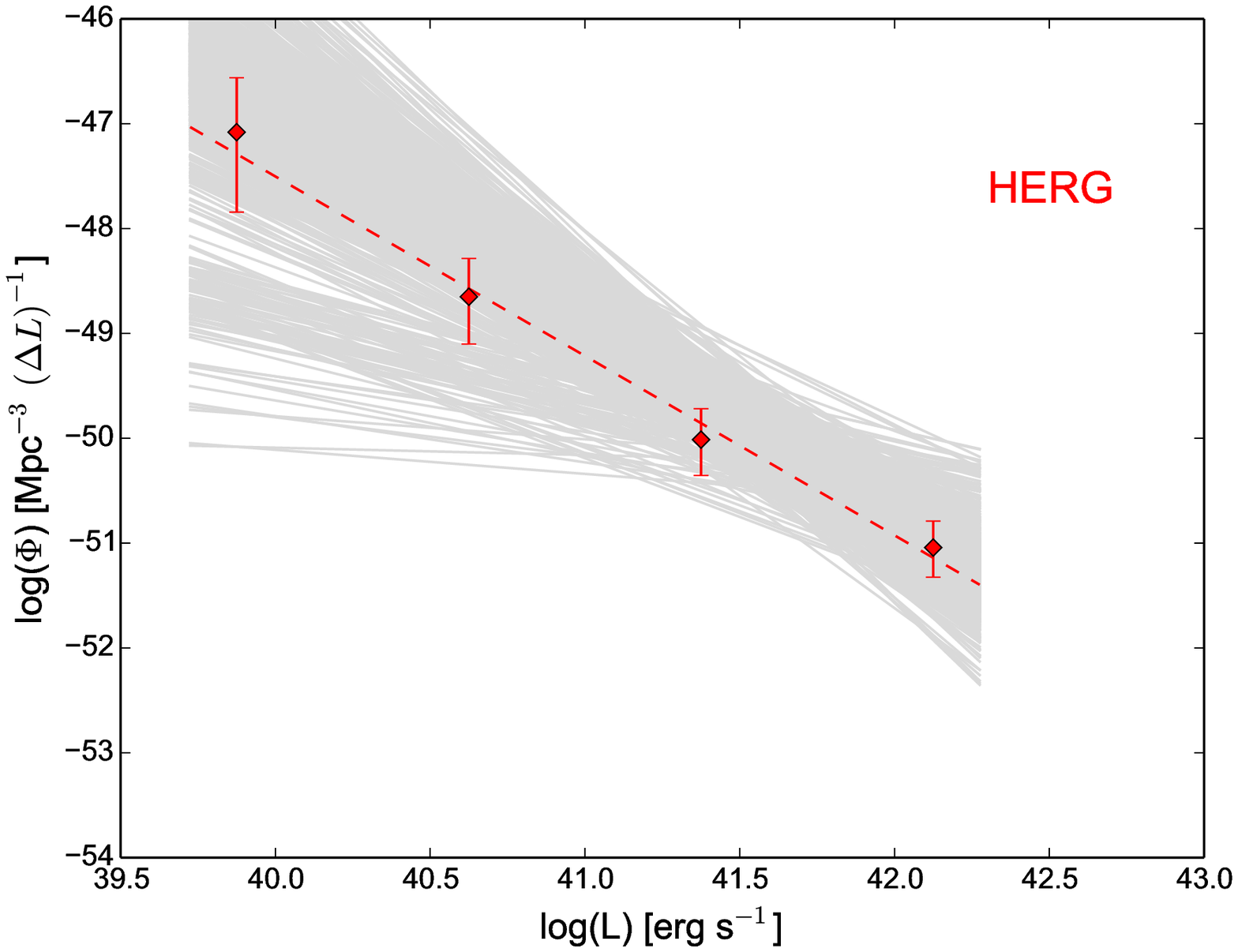}
\caption{Monochromatic radio luminosity functions at 1.4 GHz. Top panel: F-NLS1s; bottom panel: HERGs. Dashed lines indicate the single power law that is best fit. The light gray lines indicate the simulated luminosity functions obtained via Monte Carlo method, as described in the text.}
\label{fig:lf2}
\label{multiplot}
\end{figure}
\section{Luminosity functions}
\begin{table*}[t!]
\centering
\caption{Parameters of the luminosity functions. The LF of F-NLS1s is showed in two different ways: with or without 1 source with unknown spectral index. The former is indicated with an asterisk. }
\centering
\begin{tabular}{l c c c c c c c c }
\hline\hline
Sample & Mod. & $\log$ L$_{1}$ & $\log$ L$_{2}$ & $\log$ L$_b$ & $\log\Phi_b$ & $\log{K}$ & $\alpha$ & $\beta$ \\
\hline
F-NLS1 & PL & 40.25 & 42.75 & $-$ & $-$ & -1.91$\pm$5.90 & -1.17$\pm$0.14$\pm$0.48 & $-$ \\
F-NLS1* & PL & 40.25 & 42.75 & $-$ & $-$ & -1.60$\pm$5.90 & -1.18$\pm$0.14$\pm$0.48 & $-$ \\
HERG & PL & 39.75 & 42.25 & $-$ & $-$ & 20.96$\pm$5.68 & -1.71$\pm$0.14$\pm$0.41 & $-$ \\
FSRQ & BPL & 42.25 & 43.75 & 43.32$\pm$0.47 & -53.10$\pm$1.34 & $-$  & 1.61$\pm$0.98 & 4.33$\pm$2.00 \\
\hline
\label{tab:lf}
\end{tabular}
\tablefoot{Columns: (1) sample; (2) function used for the best-fit. PL for power law, BPL for broken power law; (3) logarithm of minimum luminosity bin (\ergs); (4) logarithm of maximum luminosity bin (\ergs); (5) logarithm of luminosity break (\ergs); (6) logarithm of the luminosity function at the break (Mpc$^{-3}$); (7) coefficient of the power law; (8) slope of the power law (slope below the break for broken power law). The second error is that evaluated via Monte Carlo method; (9) slope above the break (for broken power law only).}
\end{table*}
\subsection{Method}
The luminosity function (LF) describes the volumetric density of sources as a function of their luminosity. For flux-limited samples the LF is computed as in \citet{Peterson}
\begin{equation}
\Phi(L) = \frac{1}{\Delta L}\frac{4\pi}{A}\sum_{L_i \in (L \pm \Delta L/2)} \frac{1}{V_{max}(L)} \; ,
\end{equation}
where $\Delta$L is the width of the luminosity bin, and A is the area of sky covered by the samples. In our cases, the area covered both by the DR5 and FIRST is $\sim$1/7 of the whole sky, while the common area between DR7 and FIRST is $\sim$1/6. \par
To compute the LF we divided the sources in bins of luminosity (L$-\Delta$L/2, L$+\Delta$L/2). In those samples that have a lower redshift limit, instead of V$_{max}$ we used the accessible volume V$_a$. We assume that the only source of uncertainty in the LF is the error on the number counts per bin, hence we assumed a Poissonian statistics. The Poissonian statistics is not symmetric for small values (N $\lesssim$ 10); to evaluate the errors in the low statistic limit we used the values from \citet{Gehrels86}. \par
Our aim was to determine the radio LF for each sample. We then calculated the luminosity at 1.4 GHz for each source from the peak flux of the FIRST survey. We performed a K-correction, using all the spectral indices we found in the literature. For the only F-NLS1 with unknown spectral index, we assumed a flat spectrum ($\alpha_\nu = 0$). We divided the luminosities in bins of 0.25 dex for the control sample, since there were enough data to fill each bin. In the other two cases, we used a binning of 0.5 dex for F-NLS1s and of 0.75 dex for HERGs. The LFs were fitted with a single power law
\begin{equation}
\Phi(L) = KL^\alpha \; ,
\end{equation}
where K is a constant and $\alpha$ the slope of the power law. In the case of FSRQs, following \citet{Padovani92}, we used a broken power law in the form
\begin{equation}
\Phi(L) = \frac{\Phi_b}{\left(L/L_b\right)^\alpha+\left(L/L_b\right)^\beta} \; ,
\end{equation}
where $\Phi_b$ is the normalization factor, $L_b$ is the break luminosity, and $\alpha$ and $\beta$ are the two slopes. In the FSRQs sample we also applied a correction for luminosity evolution to bring each source to z = 0. For this purpose, we assumed the same cosmological evolution found by \citet{Padovani92}, $\exp{(-T/\tau)}$, where $T$ is the lookback time and $\tau = 0.23$ is the timescale of evolution in units of Hubble time. Using the spectral indices we also derived the 1.4 GHz flux for each FSRQ to enable a direct comparison with the other samples. All fits were performed using the generalized least squares method. The results are shown in Fig.~\ref{fsrq_lf} and \ref{fig:lf2} and summarized in Tab.~\ref{tab:lf}. \par
To take into account the small size of F-NLS1s and HERGs samples, we decided to provide a further estimate of the errors using a Monte Carlo method. Starting from the observed luminosity functions, we generated two large samples of $\sim$3$\times$10$^5$ simulated F-NLS1s and of 2$\times10^6$ simulated HERGs, using the same selection criteria of our samples. From these larger samples, we extracted a random number  of F-NLS1s between 10 and 16, and a random number of HERGs between 7 and 13. We then evaluated on these new samples the luminosity function. This operation was repeated 1000 times. The simulated luminosity functions are plotted in light gray in Fig.~\ref{fig:lf2}. The spread in slopes obtained is also reported in Tab.~\ref{tab:lf}. \par
\begin{figure}[t!]
\centering
\includegraphics[width=\hsize]{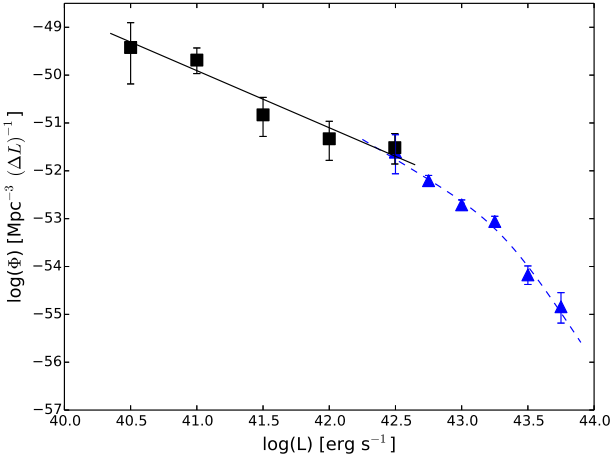} 
\caption{Monochromatic radio luminosity functions of F-NLS1s and FSRQs at 1.4 GHz. The black squares indicate the F-NLS1s data points, the blue triangles indicate the FSRQs data points. The blue dashed line represents the broken power law best fit for FSRQs, and the black solid line represents the single power law that is best fit for F-NLS1s.}
\label{fsrq_flat}
\end{figure}

Our FSRQs LF is in agreement with that obtained by \citet{Padovani92} when the cosmology they adopted is used. In F-NLS1s, the inclusion in the LF of the two sources with undetermined spectral index has a negligible impact, since neither the slope nor the coefficient of the LF are significantly affected (Tab.~\ref{tab:lf}). In both cases, the scatter is fairly high, likely because of the low statistic. The slope of HERGs is steeper than that of F-NLS1s, which indeed have a nearly flat LF. In particular, the slope of F-NLS1s is close to that of FSRQs for luminosities below the break, even if the error on this slope is large. This result becomes more evident when the LFs of F-NLS1s and FSRQs are shown together, as in Fig.~\ref{fsrq_flat}. The two LFs are very close in the region of $10^{43}$ \ergs\ and the LF of F-NLS1s appears to be an extension of that of FSRQs at lower luminosities. 
\subsection{Relativistic beaming}
In order to compare the beamed sources with their parent population, we have to take into account the effect of beaming on the LF shape. We then added the relativistic beaming to the CSS/HERGs luminosity function. This cannot be carried out analytically, as explained by \citet{Urry84} and \citet{Urry91}. We followed the procedure described by \citet{Urry84} for a single power law. In analogy with that work, we defined as $\lll$ the intrinsic luminosity, where $L$ is the observed luminosity. These two quantities are related via $L = \delta^p \lll$, where $\delta = [\Gamma(1-\beta\cos\theta)]^{-1}$ is the kinematic Doppler factor of the jet and the exponent is $p = 3 + \alpha$, where $\alpha$ is the intrinsic slope of the jet emission. The total flux emitted by the source is given by L = (1 + $f\delta^p$)$\lll_u$, where $\lll_u$ is the unbeamed luminosity and $f$ is the ratio between the jet luminosity and unbeamed luminosity. The model is then evaluated numerically via
\begin{equation}
\Phi(L) = \int \frac{K}{\beta\gamma p} f^{1/p} \lll^{\alpha-1} \left(\frac{L}{\lll}-1\right)^{-(p+1)/p} \mathrm{d}\lll \; .
\end{equation}
We used $p = 3.7$ because the typical slope of a synchrotron spectrum is $\alpha = 0.7$. We also performed our calculations for different values of $f$ (0.01 $\leq f \leq$ 1) and bulk Lorentz factor, 8 $\leq \Gamma \leq$ 15, which are values already observed in $\gamma$-ray emitting NLS1s \citep{Abdo09c, Dammando12}. To evaluate the Doppler factor we assumed the angle to vary between $0^\circ \leq \theta \leq \theta_c$, where $\theta_c$ is the critical angle for which $\delta(\Gamma,f,\theta_c) = 1$. Therefore all the sources with inclination $\theta$ appear as F-NLS1s. In the case of a simple power law, the resulting beamed LF is a broken power law. \par
\begin{figure}[t!]
\centering
\includegraphics[width=\hsize]{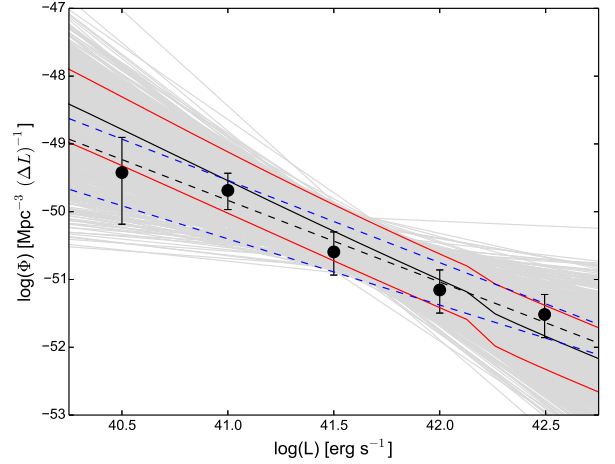} 
\caption{HERGs LF with relativistic beaming added for bulk Lorentz factor $\Gamma = 10$ and ratio $f$ = 1. Black solid line indicates the model; red solid lines indicate the maximum and minimum values for the model. Black circles show F-NLS1s data, black dashed line shows the F-NLS1s LF best-fit, and blue dashed lines indicate the maximum and minimum values for F-NLS1s LF. The light gray lines denote the simulated LFs for F-NLS1s.}
\label{beam}
\end{figure}

We derived the maximum and minimum values allowable for the data from the analytical error bars. The errors in the models of beamed LF are evaluated by refitting such maximum and minimum values. These new fits were performed using the same functions adopted for the previous fitting of the data. We then added the relativistic beaming both to the best fit, the minimum, and maximum fit. The resulting parent+beaming model is shown in Fig.~\ref{beam}. \par
We evaluated the distance between the model and our data by means of the reduced chi-squared, $\chi^2_\nu$ to test our results. The results are shown in Tab.~\ref{chi2}. We report the $\chi^2_\nu$ of the model and the lowest $\chi^2_\nu$ also considering the maximum and minimum curve. We also evaluated the model for several values of bulk Lorentz factor, to understand up to which values the model was still acceptable. We must underline that our hypothesis of constant $\Gamma$ in all sources is likely unrealistic, and that a a power law distribution of $\Gamma$ values is probably closer to reality \citep{Lister97, Liodakis15}. The inclusion of an additional parameter in the analysis is not statistically justified, however, given the large uncertainties on the observed LFs.
\par
\begin{table}[t!]
\centering
\caption{$\chi^2_\nu$ for the beaming model tested with different parameters. The star indicates that the F-NLS1s sample also included the two sources with unknown spectral index.  }
\begin{tabular}{l c c c c c}
\hline\hline
Sample & $\Gamma$ & $f$ & $\chi^2_\nu$ & $\chi^2_\nu$ (max) & $\chi^2_\nu$ (min) \\
\hline
HERG & 10 & 1.0 & 1.95 & 8.0 & 3.85 \\
HERG* & 10 & 1.0 & 1.53 & 7.79 & 3.89 \\
HERG & 8 & 1.0 & 1.97 & 8.77 & 3.49 \\
HERG* & 8 & 1.0 & 1.55 & 8.61 & 3.53 \\
HERG & 15 & 1.0 & 2.67 & 7.45 & 5.35 \\
HERG* & 15 & 1.0 & 2.25 & 7.19 & 5.38 \\
HERG & 10 & 0.5 & 2.36 & 7.59 & 4.82 \\
HERG* & 10 & 0.5 & 1.94 & 7.35 & 4.86 \\
HERG & 10 & 0.1 & 4.12 & 8.72 & 6.39 \\
HERG* & 10 & 0.1 & 3.69 & 8.51 & 6.28 \\
HERG & 10 & 0.01 & 20.02 & 29.51 & 15.16 \\
HERG* & 10 & 0.01 & 20.42 & 30.84 & 14.89 \\
\hline
\end{tabular}
\label{chi2}
\tablefoot{Columns: (1) sample; (2) bulk Lorentz factor of the jet; (3) ratio between beamed and diffuse emission from the jet $f$; (4) $\chi^2_\nu$ of the model; (5) $\chi^2_\nu$ of the maximum model; and (6) $\chi^2_\nu$ of the minimum model.}
\end{table}
As shown in Fig.~\ref{beam}, the best-fit power law for F-NLS1s and the model prediction are in good agreement, but there is a deviation at lower luminosities. In particular, the slope of the model in the region occupied by F-NLS1s is -1.52, while the slope of the measured LF is -1.17. The values of $\chi^2_\nu$ are not very close to 1, largely because of this deviation. We think that this discrepancy is due to a selection effect. Our F-NLS1s sample includes only very radio-loud NLS1s, therefore the resulting luminosity function might be underestimated in the low-luminosity region. Nevertheless, the beaming model and its errors are well included in the region where the simulated luminosity functions lies. Keeping all this in mind, the overlapping of the model with the observed function is quite satisfactory. \par
In Tab.~\ref{chi2} we report the values of $\chi^2_\nu$ calculated with the different values of $\Gamma$ and $f$. We highlight that the closest $\chi^2_\nu$ between the model and data is observed in the sample which includes the two sources with undetermined spectral index, for a ratio $f$ = 1.0 and $\Gamma = 10$, which translates in a slope of the model of -1.52. This value of $f$ is significantly higher than that observed in FSRQs, which is between $10^{-3}-10^{-2}$ \citep{Padovani92}. \par
\section{Discussion}
\subsection{Black hole mass}
The first result that must be highlighted is that the black hole mass distribution of CSS/HERGs is similar to that of F-NLS1s with typical values between 10$^7$ and 10$^8$ M$_\odot$. Also the Eddington ratio is high, comparable to that of typical NLS1s, both radio loud and quiet. This result is expected if NLS1s and CSS/HERGs both have a radiatively efficient accretion mechanism, that is similar to that of FSRQs. \par
The K-S test revealed that the distributions of these quantities, both mass and Eddington ratio, in F-NLS1s and CSS/HERGs might be drawn from the same population. The most obvious interpretation of this result is that CSS/HERGs might actually be misaligned F-NLS1s. Of course this result is obtained for very small samples, so it must be taken with some caution. In particular the masses of CSS/HERGs, as they are derived using forbidden lines, must be considered only an upper limit. If the narrow-line region is perturbed because of interaction with the relativistic jet, the FWHM is indeed higher and leads to an overestimate of the mass. Nevertheless our findings are in good agreement with those of previous works, where the similarity between CSS/HERGs and NLS1s was already pointed out. For example \citet{Wu09} found that a large number of CSS/GPS have a black hole mass between 10$^{7.5}$ and 10$^8$ M$_\odot$, and the same conclusion was obtained by \citet{Son12} again on CSS, both for HERGs and LERGs. Moreover the sample by \citet{Foschini15}, of which ours is a subset, revealed that on average F-NLS1s also have black hole mass between 10$^7$ and 10$^8$ M$_\odot$. All these results seem to support our hypothesis. 
\subsection{Evolutionary picture}
The V/V$_{max}$ test shows that both F-NLS1s and CSS/HERGs have no significant luminosity and/or density evolution up to z $=$ 0.6. Instead, FSRQs instead show a strong luminosity evolution, but the sample is extended to much larger distances. An interesting result we found is shown in Fig.~\ref{fsrq_flat}, and it might point out that FSRQs and F-NLS1s are strictly connected to each other. F-NLS1s were suggested to be the low-mass tail of $\gamma$-ray emitting AGN, and in particular of FSRQs \citep[][and references therein]{Foschini15}. Since the black hole mass and the jet power are connected \citep{Heinz03}, the lower radio luminosity and jet power of F-NLS1s might be a consequence of the lower black hole mass. Therefore it is expected that F-NLS1s would be the low-luminosity tail of FSRQs LF, as we indeed observe. Of course, there might be some low-luminosity FSRQs that cannot be classified as NLS1s. The criteria for NLS1s classification is indeed based mainly on the H$\beta$ width, which is not just a function of the black hole mass. Therefore not all low mass FSRQs can be classified as NLS1s, even if their black hole mass and radio emission are comparable. \par
An explanation for the low mass is the young scenario of NLS1s. If this is true, F-NLS1s might be the young counterpart of FSRQs in which the nuclear activity started only recently and in which the black hole (and possibly the host galaxy) is still (co-)evolving. 
A similar picture was already suggested for CSS sources years ago \citep{Readhead96, Fanti95, Odea97}. These likely young radio sources are thought to be an evolutionary phase that is going to evolve into the giant double sources. In particular, \citet{Kunert10b} also took into account the optical division into HERG and LERG, finding that the CSS/HERGs sources are likely going to evolve into FR$_{HERG}$. Recently, \citet{Giommi12} suggested that the two blazar classes, and hence their parent population, should be divided according to their low or high ionization, and that all the other classifications are physically irrelevant. If this is true, FSRQs can be identified as beamed HERGs, and F-NLS1s, which might be young FSRQs, should be the beamed version of young HERGs, so CSS/HERGs. In summary, the evolutionary picture for beamed sources might be simply F-NLS1 $\rightarrow$ FSRQ, and for their parent population CSS/HERGs $\rightarrow$ FR$_{HERG}$. An evolutionary connection between F-NLS1s and FSRQs is then possible, where the former are still growing to become the latter. This hypothesis finds further support in our Fig.~\ref{beam}. When the relativistic beaming is added using the typical bulk Lorentz factor of $\gamma$-ray emitting NLS1s, CSS/HERGs LF reproduces well the data. Even if at low luminosities the model predicts a larger number of F-NLS1s that we do not observe, we think that this discrepancy might only be due to the selection criterion of our NLS1s sample. Keeping this caveat in mind, the model seems to indicate that CSS/HERGs might be good parent candidates. \par
In young radio sources such as CSS, the jet activity might be intermittent, and several outburst episodes might be induced by pressure radiation instabilities in the accretion disk, with a timescale of 10$^2$-10$^5$ years \citep{Czerny09, Wu09b}. A similar, strong variability is also observed in RLNLS1s \citep{Foschini12a, Foschini15}, providing further confirmation for this unified model. If CSS/HERGs are parent sources, the origin for these activity/inactivity phases in F-NLS1s might be the same. The radiation pressure instability is indeed one of the hypotheses that can account for the nonthermal emission and extended structures observed in some radio-quiet NLS1s \citep{Doi12}, where the jet activity phase might have lasted for only a few years thereby leaving the observed structures \citep{Ghisellini04}. \par
The inclusion of CSS/HERGs in the parent population of F-NLS1s might moreover rule out the vast majority of RQNLS1s as parent candidates. In fact, since CSS/HERGs display lobes that are already developed \citep{Orienti15}, this means that the extended radio emission form even in very young ages, which in turn implies the rejection of the radio-quiet hypothesis. This conclusion is in agreement with the results of \citet{Berton15b}, where the observed differences in narrow-line region properties points in the same direction. \par
Another aspect to consider is the role that S-NLS1s can play in this scenario. These sources are likely misaligned F-NLS1s \citep{Berton15a}, therefore for this picture to be coherent they should also be part of the larger class of CSS/HERGs. The sample we used in this work unfortunately does not include any type 1 AGN, so our data can reveal nothing on this issue, but this topic has already been investigated in the literature, particularly in recent years. Several authors found that at least some S-NLS1s can indeed be classified as CSS/HERGs \citep{Caccianiga14, Komossa15, Schulz15}. In particular the extended survey by \citet{Gu15} showed that the radio morphology of almost each one of their S-NLS1s closely recalls that of CSS. These results are therefore in agreement with our hypothesis, and seem to favor the scenario in which CSS/HERGs are the largest class of F-NLS1s parent sources. It is also reasonable that S-NLS1s are objects observed at intermediate angles between F-NLS1s and obscured (type 2) CSS/HERGs. \par
Still, it is not clear whether all type 1 CSS/HERGs are NLS1s. Few CSS/HERGs have lines with a FWHM(H$\beta$) $>$ 2000 \kms, and cannot be classified as NLS1s. It is then possible that the unification between CSS/HERGs and NLS1s is only in a statistical sense, that is CSS/HERGs and NLS1s are on average the same population, but with few exceptions likely connected to the NLS1s definition. If a more physical classification was used, such as black hole mass or Eddington ratio, the unification between these sources would show fewer outliers.\par In any case, it is also possible that the BLR geometry has some impact on these outliers. If a flattened component in the BLR is present, sources with a large inclination should appear as broad-line AGN and not classifiable as NLS1s. The presence of some relatively high mass type 1 sources in a CSS/HERGs sample might then provide a clue to the BLR geometry. Anyway it is worth noting that in the sample of CSS by \citet{Son12}, the type 1 AGN have a BH mass always below 5$\times$10$^8$ M$_\odot$, and an average value of 8.9$\times$10$^7$ M$_\odot$. If these sources had a flattened BLR and were randomly oriented, some of these sources would show a much larger mass. Instead all the values are in good agreement with those of F-NLS1s, so they do not seem to have a flattened component in the BLR. In any case, a deeper study on a larger sample is necessary to better address this problem. \par

\subsection{Host galaxy}
A possible objection to the identification of CSS/HERGs as the parent population of F-NLS1s is that their host galaxy might be different. In particular CSS, like many radio-loud AGN, are usually thought to be hosted by elliptical galaxies \citep{Best05, Orienti15} and triggered by merging activity \citep{Holt09}. Instead, NLS1s are generally believed to be hosted by spiral galaxies \citep{Crenshaw03} with a pseudobulge formed via secular evolution \citep{Orbandexivry11, Mathur12}. Nevertheless, the K-S test we performed, along with other studies, seems to draw a more complicated picture. \par
CSS/HERGs have a black hole mass distribution closer to that of disk RGs than to that of elliptical RGs, typically showing lower black hole mass. This might be because of the young age of these sources, so perhaps the black hole is still growing to reach the mass value of typical elliptical. Nevertheless, the black hole mass is directly connected with the bulge dynamics, particularly with its stellar velocity dispersion \citep{Ferrarese00}. So in principle a relatively low mass black hole should be hosted in the small bulge of a disk galaxy, rather than in a more massive elliptical. Therefore it is possible that CSS/HERGs are also hosted in disk galaxies, as NLS1s. An example of a disk host for a powerful CSS was found by \citet{Morganti11}. Moreover in a very large sample of AGN, \citet{Best12} found that HERGs host galaxies have different properties than those of LERGs and, in particular, these authors found that they are bluer with lower mass, lower 4000 \AA{} break, and a higher star formation rate. Such characteristics are reminiscent of those of a disk or star-forming galaxy. \par
Finally, we must underline that not much is known even about the host galaxies of F-NLS1s, mainly because of their high redshift. Few studies were performed on the closest F-NLS1, 1H 0323+342, and seem to suggest the presence of a disk and possibly of a pseudobulge \citep{Anton08,Hamilton12,Leontavares14}. The WISE colors of F-NLS1s suggest that their host galaxy has a higher rate of ongoing star formation \citep{Caccianiga15}, a feature that also reveals an ``active" host and is likely different from giant ``passive'' elliptical galaxies. However, further studies on the host galaxy morphology are necessary to characterize F-NLS1s as a population.

\section{Summary}
In this paper we investigated the relation that exists between CSS sources with an HERG optical spectrum and F-NLS1s. Our aim was to understand whether the CSS/HERGs class can be part of the parent population of F-NLS1s. To do this, we analyzed the only two statistically complete samples of CSS/HERG and F-NLS1 available so far. First we calculated the black hole mass and Eddington ratio by means of the optical spectrum, and then we studied their radio luminosity functions along with that of a control sample of FSRQs. \par
The black hole masses are tipically between 10$^{7.5}$ and 10$^8$ M$_\odot$ in both samples, and the Eddington ratio is around $\sim$0.1. We performed a Kolmogorov-Smirnov test on the samples to compare their black hole mass distributions. Our results, in agreement with previous studies in the literature, seem to confirm that the two distributions might be drawn from the same population and, hence, that CSS/HERGs are good candidates for parent sources. \par
The luminosity functions seem to support the same scenario. A first result is that F-NLS1s might be the low-luminosity (and low-mass) tail of FSRQs, confirming the results of \citet{Abdo09a} and \citet{Foschini15}. The addition of relativistic beaming to CSS/HERGs luminosity function revealed that CSS/HERGs might actually be F-NLS1s with the jet viewed at large angle, and thus belonging to the parent population. In this framework, RLNLS1s with a steep radio spectrum are sources observed at intermediate angle between F-NLS1s and CSS/HERGs with a type 2 (absorbed) optical spectrum. \par
Our results also seem to be consistent with an evolutionary picture in which F-NLS1s and CSS/HERGs are the young and still growing phases of FSRQs and FR$_{HERG}$, respectively. A more detailed study is required on a larger sample of sources. In particular, new spectral indices are necessary to effectively compare CSS/HERGs and F-NLS1s. New large surveys at different frequencies, such as VLASS, might be helpful to improve our knowledge about these sources. Also SKA, with its unprecedented sensitivity, will likely provide an incredible amount of information to greatly deepen our understanding of RLNLS1s \citep{Berton16a}. 

\begin{acknowledgements}
We thank M.~L. Lister and R. Antonucci for helpful suggestions and discussions. We thank our referee Vasiliki Pavlidou for helpful comments that improved the quality of the paper. GT is partially supported by the PRIN-INAF 2014 with the project Transient Universe: unveiling new types of stellar explosions with PESSTO. The research leading to these results has received funding from the European Research Council under the European Union's Seventh Framework Programme (FP7/2007-2013)/ERC Grant agreement n$^{\rm o}$ [291222] (PI : S. J. Smartt) and STFC grants ST/L000709/1. This research has made use of the NASA/IPAC Extragalactic Database (NED), which is operated by the Jet Propulsion Laboratory, California Institute of Technology, under contract with the National Aeronautics and  Space Administration. Funding for the Sloan Digital Sky Survey has been provided by the Alfred P. Sloan Foundation and the U.S. Department of Energy Office of Science. The SDSS web site is \texttt{http://www.sdss.org}. SDSS-III is managed by the Astrophysical Research Consortium for the Participating Institutions of the SDSS-III Collaboration including the University of Arizona, the Brazilian Participation Group, Brookhaven National Laboratory, Carnegie Mellon University, University of Florida, the French Participation Group, the German Participation Group, Harvard University, the Instituto de Astrofisica de Canarias, the Michigan State/Notre Dame/JINA Participation Group, Johns Hopkins University, Lawrence Berkeley National Laboratory, Max Planck Institute for Astrophysics, Max Planck Institute for Extraterrestrial Physics, New Mexico State University, University of Portsmouth, Princeton University, the Spanish Participation Group, University of Tokyo, University of Utah, Vanderbilt University, University of Virginia, University of Washington, and Yale University.
\end{acknowledgements}

\bibliographystyle{aa}
\bibliography{/home/marco/Scrivania/Tesi_dottorato/biblio}

\begin{thebibliography}{81}
\expandafter\ifx\csname natexlab\endcsname\relax\def\natexlab#1{#1}\fi

\bibitem[{{Abdo} {et~al.}(2009{\natexlab{a}}){Abdo}, {Ackermann}, {Ajello},
  {Axelsson}, {Baldini}, {Ballet}, {Barbiellini}, {Bastieri}, {Battelino}, \&
  {Baughman}}]{Abdo09a}
{Abdo}, A.~A., {Ackermann}, M., {Ajello}, M., {et~al.} 2009{\natexlab{a}},
  \apj, 699, 976

\bibitem[{{Abdo} {et~al.}(2009{\natexlab{b}}){Abdo}, {Ackermann}, {Ajello},
  {Axelsson}, {Baldini}, {Ballet}, {Barbiellini}, {Bastieri}, {Baughman},
  {Bechtol}, \& et~al.}]{Abdo09b}
{Abdo}, A.~A., {Ackermann}, M., {Ajello}, M., {et~al.} 2009{\natexlab{b}},
  \apj, 707, 727

\bibitem[{{Abdo} {et~al.}(2009{\natexlab{c}}){Abdo}, {Ackermann}, {Ajello},
  {Baldini}, {Ballet}, {Barbiellini}, {Bastieri}, {Bechtol}, {Bellazzini}, \&
  {Berenji}}]{Abdo09c}
{Abdo}, A.~A., {Ackermann}, M., {Ajello}, M., {et~al.} 2009{\natexlab{c}},
  \apjl, 707, L142

\bibitem[{{Ant{\'o}n} {et~al.}(2008){Ant{\'o}n}, {Browne}, \&
  {March{\~a}}}]{Anton08}
{Ant{\'o}n}, S., {Browne}, I.~W.~A., \& {March{\~a}}, M.~J. 2008, \aap, 490,
  583

\bibitem[{{Antonucci}(2002)}]{Antonucci02}
{Antonucci}, R. 2002, in Astrophysical Spectropolarimetry, ed.
  J.~{Trujillo-Bueno}, F.~{Moreno-Insertis}, \& F.~{S{\'a}nchez}, 151--175

\bibitem[{{Avni} \& {Bahcall}(1980)}]{Avni80}
{Avni}, Y. \& {Bahcall}, J.~N. 1980, \apj, 235, 694

\bibitem[{{Berton} {et~al.}(2016){Berton}, {Foschini}, {Caccianiga},
  {Richards}, {Ciroi}, {Congiu}, {Cracco}, {La Mura}, {Marafatto}, \&
  {Rafanelli}}]{Berton16a}
{Berton}, M., {Foschini}, L., {Caccianiga}, A., {et~al.} 2016, ArXiv e-prints
  [\eprint[arXiv]{1601.05791}]

\bibitem[{{Berton} {et~al.}(2015{\natexlab{a}}){Berton}, {Foschini}, {Ciroi},
  {Cracco}, {La Mura}, {Di Mille}, \& {Rafanelli}}]{Berton15b}
{Berton}, M., {Foschini}, L., {Ciroi}, S., {et~al.} 2015{\natexlab{a}}, ArXiv
  e-prints [\eprint[arXiv]{1506.05800}]

\bibitem[{{Berton} {et~al.}(2015{\natexlab{b}}){Berton}, {Foschini}, {Ciroi},
  {Cracco}, {La Mura}, {Lister}, {Mathur}, {Peterson}, {Richards}, \&
  {Rafanelli}}]{Berton15a}
{Berton}, M., {Foschini}, L., {Ciroi}, S., {et~al.} 2015{\natexlab{b}}, \aap,
  578, A28

\bibitem[{{Best} \& {Heckman}(2012)}]{Best12}
{Best}, P.~N. \& {Heckman}, T.~M. 2012, \mnras, 421, 1569

\bibitem[{{Best} {et~al.}(2005){Best}, {Kauffmann}, {Heckman}, {Brinchmann},
  {Charlot}, {Ivezi{\'c}}, \& {White}}]{Best05}
{Best}, P.~N., {Kauffmann}, G., {Heckman}, T.~M., {et~al.} 2005, \mnras, 362,
  25

\bibitem[{{Boroson} \& {Green}(1992)}]{Boroson92}
{Boroson}, T.~A. \& {Green}, R.~F. 1992, \apjs, 80, 109

\bibitem[{{Caccianiga} {et~al.}(2014){Caccianiga}, {Ant{\'o}n}, {Ballo},
  {Dallacasa}, {Della Ceca}, {Fanali}, {Foschini}, {Hamilton}, {Kraus},
  {Maccacaro}, {Mack}, {March{\~a}}, {Paulino-Afonso}, {Sani}, \&
  {Severgnini}}]{Caccianiga14}
{Caccianiga}, A., {Ant{\'o}n}, S., {Ballo}, L., {et~al.} 2014, \mnras, 441, 172

\bibitem[{{Caccianiga} {et~al.}(2015){Caccianiga}, {Ant{\'o}n}, {Ballo},
  {Foschini}, {Maccacaro}, {Della Ceca}, {Severgnini}, {March{\~a}}, {Mateos},
  \& {Sani}}]{Caccianiga15}
{Caccianiga}, A., {Ant{\'o}n}, S., {Ballo}, L., {et~al.} 2015, \mnras, 451,
  1795

\bibitem[{{Crenshaw} {et~al.}(2003){Crenshaw}, {Kraemer}, \&
  {Gabel}}]{Crenshaw03}
{Crenshaw}, D.~M., {Kraemer}, S.~B., \& {Gabel}, J.~R. 2003, \aj, 126, 1690

\bibitem[{{Czerny} {et~al.}(2009){Czerny}, {Siemiginowska}, {Janiuk},
  {Nikiel-Wroczy{\'n}ski}, \& {Stawarz}}]{Czerny09}
{Czerny}, B., {Siemiginowska}, A., {Janiuk}, A., {Nikiel-Wroczy{\'n}ski}, B.,
  \& {Stawarz}, {\L}. 2009, \apj, 698, 840

\bibitem[{{D'Ammando} {et~al.}(2012){D'Ammando}, {Orienti}, {Finke}, {Raiteri},
  {Angelakis}, {Fuhrmann}, {Giroletti}, {Hovatta}, {Max-Moerbeck}, {Perkins},
  {Readhead}, {Richards}, {Stawarz}, \& {Donato}}]{Dammando12}
{D'Ammando}, F., {Orienti}, M., {Finke}, J., {et~al.} 2012, \mnras, 426, 317

\bibitem[{{de Bruyn} \& {Wilson}(1978)}]{debruyn78}
{de Bruyn}, A.~G. \& {Wilson}, A.~S. 1978, \aap, 64, 433

\bibitem[{{Decarli} {et~al.}(2008){Decarli}, {Dotti}, {Fontana}, \&
  {Haardt}}]{Decarli08}
{Decarli}, R., {Dotti}, M., {Fontana}, M., \& {Haardt}, F. 2008, \mnras, 386,
  L15

\bibitem[{{Doi} {et~al.}(2011){Doi}, {Asada}, \& {Nagai}}]{Doi11}
{Doi}, A., {Asada}, K., \& {Nagai}, H. 2011, \apj, 738, 126

\bibitem[{{Doi} {et~al.}(2012){Doi}, {Nagira}, {Kawakatu}, {Kino}, {Nagai}, \&
  {Asada}}]{Doi12}
{Doi}, A., {Nagira}, H., {Kawakatu}, N., {et~al.} 2012, \apj, 760, 41

\bibitem[{{Fanti} {et~al.}(1995){Fanti}, {Fanti}, {Dallacasa}, {Schilizzi},
  {Spencer}, \& {Stanghellini}}]{Fanti95}
{Fanti}, C., {Fanti}, R., {Dallacasa}, D., {et~al.} 1995, \aap, 302, 317

\bibitem[{{Ferrarese} \& {Merritt}(2000)}]{Ferrarese00}
{Ferrarese}, L. \& {Merritt}, D. 2000, \apjl, 539, L9

\bibitem[{{Foschini}(2011)}]{Foschini11}
{Foschini}, L. 2011, in Narrow-Line Seyfert 1 Galaxies and their Place in the
  Universe, Proc. of Science, Vol. NLS1, id. 24

\bibitem[{{Foschini}(2012)}]{Foschini12}
{Foschini}, L. 2012, in Proceedings of Nuclei of Seyfert galaxies and QSOs -
  Central engine \& conditions of star formation, Proc. of Science, Vol.
  Seyfert 2012, id. 10

\bibitem[{{Foschini} {et~al.}(2012){Foschini}, {Angelakis}, {Fuhrmann},
  {Ghisellini}, {Hovatta}, {Lahteenmaki}, {Lister}, {Braito}, {Gallo},
  {Hamilton}, {Kino}, {Komossa}, {Pushkarev}, {Thompson}, {Tibolla},
  {Tramacere}, {Carrami{\~n}ana}, {Carrasco}, {Falcone}, {Giroletti}, {Grupe},
  {Kovalev}, {Krichbaum}, {Max-Moerbeck}, {Nestoras}, {Pearson}, {Porras},
  {Readhead}, {Recillas}, {Richards}, {Riquelme}, {Sievers}, {Tammi},
  {Tornikoski}, {Ungerechts}, {Zensus}, {Celotti}, {Bonnoli}, {Doi},
  {Maraschi}, {Tagliaferri}, \& {Tavecchio}}]{Foschini12a}
{Foschini}, L., {Angelakis}, E., {Fuhrmann}, L., {et~al.} 2012, \aap, 548, A106

\bibitem[{{Foschini} {et~al.}(2015){Foschini}, {Berton}, {Caccianiga}, {Ciroi},
  {Cracco}, {Peterson}, {Angelakis}, {Braito}, {Fuhrmann}, {Gallo}, {Grupe},
  {J{\"a}rvel{\"a}}, {Kaufmann}, {Komossa}, {Kovalev}, {L{\"a}hteenm{\"a}ki},
  {Lisakov}, {Lister}, {Mathur}, {Richards}, {Romano}, {Sievers},
  {Tagliaferri}, {Tammi}, {Tibolla}, {Tornikoski}, {Vercellone}, {La Mura},
  {Maraschi}, \& {Rafanelli}}]{Foschini15}
{Foschini}, L., {Berton}, M., {Caccianiga}, A., {et~al.} 2015, \aap, 575, A13

\bibitem[{{Foschini} {et~al.}(2010){Foschini}, {Fermi/Lat Collaboration},
  {Ghisellini}, {Maraschi}, {Tavecchio}, \& {Angelakis}}]{Foschini10}
{Foschini}, L., {Fermi/Lat Collaboration}, {Ghisellini}, G., {et~al.} 2010, in
  Astronomical Society of the Pacific Conference Series, Vol. 427, Accretion
  and Ejection in AGN: a Global View, ed. L.~{Maraschi}, G.~{Ghisellini},
  R.~{Della Ceca}, \& F.~{Tavecchio}, 243--248

\bibitem[{{Gallo} {et~al.}(2006){Gallo}, {Edwards}, {Ferrero}, {Kataoka},
  {Lewis}, {Ellingsen}, {Misanovic}, {Welsh}, {Whiting}, {Boller}, {Brinkmann},
  {Greenhill}, \& {Oshlack}}]{Gallo06a}
{Gallo}, L.~C., {Edwards}, P.~G., {Ferrero}, E., {et~al.} 2006, \mnras, 370,
  245

\bibitem[{{Gehrels}(1986)}]{Gehrels86}
{Gehrels}, N. 1986, \apj, 303, 336

\bibitem[{{Ghisellini} {et~al.}(2004){Ghisellini}, {Haardt}, \&
  {Matt}}]{Ghisellini04}
{Ghisellini}, G., {Haardt}, F., \& {Matt}, G. 2004, \aap, 413, 535

\bibitem[{{Giommi} {et~al.}(2012){Giommi}, {Padovani}, {Polenta}, {Turriziani},
  {D'Elia}, \& {Piranomonte}}]{Giommi12}
{Giommi}, P., {Padovani}, P., {Polenta}, G., {et~al.} 2012, \mnras, 420, 2899

\bibitem[{{Goodrich}(1989)}]{Goodrich89}
{Goodrich}, R.~W. 1989, \apj, 342, 224

\bibitem[{{Greene} \& {Ho}(2005)}]{Greene05}
{Greene}, J.~E. \& {Ho}, L.~C. 2005, \apj, 627, 721

\bibitem[{{Grupe}(2000)}]{Grupe00}
{Grupe}, D. 2000, New Astron. Rev., 44, 455

\bibitem[{{Gu} {et~al.}(2015){Gu}, {Chen}, {Komossa}, {Yuan}, {Shen}, {Wajima},
  {Zhou}, \& {Zensus}}]{Gu15}
{Gu}, M., {Chen}, Y., {Komossa}, S., {et~al.} 2015, \apjs, 221, 3

\bibitem[{{Hamilton} \& {Foschini}(2012)}]{Hamilton12}
{Hamilton}, T.~S. \& {Foschini}, L. 2012, AAS, 220

\bibitem[{{Hardcastle} {et~al.}(2007){Hardcastle}, {Evans}, \&
  {Croston}}]{Hardcastle07}
{Hardcastle}, M.~J., {Evans}, D.~A., \& {Croston}, J.~H. 2007, \mnras, 376,
  1849

\bibitem[{{Heinz} \& {Sunyaev}(2003)}]{Heinz03}
{Heinz}, S. \& {Sunyaev}, R.~A. 2003, \mnras, 343, L59

\bibitem[{{Holt}(2009)}]{Holt09}
{Holt}, J. 2009, Astronomische Nachrichten, 330, 226

\bibitem[{{Kollgaard} {et~al.}(1992){Kollgaard}, {Wardle}, {Roberts}, \&
  {Gabuzda}}]{Kollgaard92}
{Kollgaard}, R.~I., {Wardle}, J.~F.~C., {Roberts}, D.~H., \& {Gabuzda}, D.~C.
  1992, \aj, 104, 1687

\bibitem[{{Komatsu} {et~al.}(2011){Komatsu}, {Smith}, {Dunkley}, {Bennett},
  {Gold}, {Hinshaw}, {Jarosik}, {Larson}, {Nolta}, {Page}, {Spergel},
  {Halpern}, {Hill}, {Kogut}, {Limon}, {Meyer}, {Odegard}, {Tucker}, {Weiland},
  {Wollack}, \& {Wright}}]{Komatsu11}
{Komatsu}, E., {Smith}, K.~M., {Dunkley}, J., {et~al.} 2011, \apjs, 192, 18

\bibitem[{{Komossa} {et~al.}(2006){Komossa}, {Voges}, {Xu}, {Mathur}, {Adorf},
  {Lemson}, {Duschl}, \& {Grupe}}]{Komossa06}
{Komossa}, S., {Voges}, W., {Xu}, D., {et~al.} 2006, \aj, 132, 531

\bibitem[{{Komossa} {et~al.}(2015){Komossa}, {Xu}, {Fuhrmann}, {Grupe}, {Yao},
  {Fan}, {Myserlis}, {Angelakis}, {Karamanavis}, {Yuan}, \&
  {Zensus}}]{Komossa15}
{Komossa}, S., {Xu}, D., {Fuhrmann}, L., {et~al.} 2015, \aap, 574, A121

\bibitem[{{Kunert-Bajraszewska} {et~al.}(2010){Kunert-Bajraszewska},
  {Gawro{\'n}ski}, {Labiano}, \& {Siemiginowska}}]{Kunert10a}
{Kunert-Bajraszewska}, M., {Gawro{\'n}ski}, M.~P., {Labiano}, A., \&
  {Siemiginowska}, A. 2010, \mnras, 408, 2261

\bibitem[{{Kunert-Bajraszewska} \& {Labiano}(2010)}]{Kunert10b}
{Kunert-Bajraszewska}, M. \& {Labiano}, A. 2010, \mnras, 408, 2279

\bibitem[{{Laing} {et~al.}(1994){Laing}, {Jenkins}, {Wall}, \&
  {Unger}}]{Laing94}
{Laing}, R.~A., {Jenkins}, C.~R., {Wall}, J.~V., \& {Unger}, S.~W. 1994, in
  Astronomical Society of the Pacific Conference Series, Vol.~54, The Physics
  of Active Galaxies, ed. G.~V. {Bicknell}, M.~A. {Dopita}, \& P.~J. {Quinn},
  201

\bibitem[{{Le{\'o}n Tavares} {et~al.}(2014){Le{\'o}n Tavares}, {Kotilainen},
  {Chavushyan}, {A{\~n}orve}, {Puerari}, {Cruz-Gonz{\'a}lez},
  {Pati{\~n}o-Alvarez}, {Ant{\'o}n}, {Carrami{\~n}ana}, {Carrasco}, {Guichard},
  {Karhunen}, {Olgu{\'{\i}}n-Iglesias}, {Sanghvi}, \& {Valdes}}]{Leontavares14}
{Le{\'o}n Tavares}, J., {Kotilainen}, J., {Chavushyan}, V., {et~al.} 2014,
  \apj, 795, 58

\bibitem[{{Liodakis} \& {Pavlidou}(2015)}]{Liodakis15}
{Liodakis}, I. \& {Pavlidou}, V. 2015, \mnras, 454, 1767

\bibitem[{{Lister} \& {Marscher}(1997)}]{Lister97}
{Lister}, M.~L. \& {Marscher}, A.~P. 1997, \apj, 476, 572

\bibitem[{{Marziani} {et~al.}(2003){Marziani}, {Zamanov}, {Sulentic}, \&
  {Calvani}}]{Marziani03}
{Marziani}, P., {Zamanov}, R.~K., {Sulentic}, J.~W., \& {Calvani}, M. 2003,
  \mnras, 345, 1133

\bibitem[{{Mathur}(2000)}]{Mathur00}
{Mathur}, S. 2000, \mnras, 314, L17

\bibitem[{{Mathur} {et~al.}(2012){Mathur}, {Fields}, {Peterson}, \&
  {Grupe}}]{Mathur12}
{Mathur}, S., {Fields}, D., {Peterson}, B.~M., \& {Grupe}, D. 2012, \apj, 754,
  146

\bibitem[{{Morganti} {et~al.}(2011){Morganti}, {Holt}, {Tadhunter}, {Ramos
  Almeida}, {Dicken}, {Inskip}, {Oosterloo}, \& {Tzioumis}}]{Morganti11}
{Morganti}, R., {Holt}, J., {Tadhunter}, C., {et~al.} 2011, \aap, 535, A97

\bibitem[{{Morganti} {et~al.}(2015){Morganti}, {Oosterloo}, {Oonk},
  {Frieswijk}, \& {Tadhunter}}]{Morganti15}
{Morganti}, R., {Oosterloo}, T., {Oonk}, J.~B.~R., {Frieswijk}, W., \&
  {Tadhunter}, C. 2015, \aap, 580, A1

\bibitem[{{Murgia} {et~al.}(1999){Murgia}, {Fanti}, {Fanti}, {Gregorini},
  {Klein}, {Mack}, \& {Vigotti}}]{Murgia99}
{Murgia}, M., {Fanti}, C., {Fanti}, R., {et~al.} 1999, \aap, 345, 769

\bibitem[{{O'Dea}(1998)}]{Odea98}
{O'Dea}, C.~P. 1998, \pasp, 110, 493

\bibitem[{{O'Dea} \& {Baum}(1997)}]{Odea97}
{O'Dea}, C.~P. \& {Baum}, S.~A. 1997, \aj, 113, 148

\bibitem[{{Orban de Xivry} {et~al.}(2011){Orban de Xivry}, {Davies},
  {Schartmann}, {Komossa}, {Marconi}, {Hicks}, {Engel}, \&
  {Tacconi}}]{Orbandexivry11}
{Orban de Xivry}, G., {Davies}, R., {Schartmann}, M., {et~al.} 2011, \mnras,
  417, 2721

\bibitem[{{Orienti}(2015)}]{Orienti15}
{Orienti}, M. 2015, ArXiv e-prints [\eprint[arXiv]{1511.00436}]

\bibitem[{{Oshlack} {et~al.}(2001){Oshlack}, {Webster}, \&
  {Whiting}}]{Oshlack01}
{Oshlack}, A.~Y.~K.~N., {Webster}, R.~L., \& {Whiting}, M.~T. 2001, \apj, 558,
  578

\bibitem[{{Osterbrock} \& {Pogge}(1985)}]{Osterbrock85}
{Osterbrock}, D.~E. \& {Pogge}, R.~W. 1985, \apj, 297, 166

\bibitem[{{Osterbrock} \& {Pogge}(1987)}]{Osterbrock87}
{Osterbrock}, D.~E. \& {Pogge}, R.~W. 1987, \apj, 323, 108

\bibitem[{{Owsianik} \& {Conway}(1998)}]{Owsianik98}
{Owsianik}, I. \& {Conway}, J.~E. 1998, \aap, 337, 69

\bibitem[{{Padovani} \& {Urry}(1992)}]{Padovani92}
{Padovani}, P. \& {Urry}, C.~M. 1992, \apj, 387, 449

\bibitem[{{Peterson}(1997)}]{Peterson}
{Peterson}, B.~M. 1997, {An Introduction to Active Galactic Nuclei}

\bibitem[{{Readhead} {et~al.}(1996){Readhead}, {Taylor}, {Pearson}, \&
  {Wilkinson}}]{Readhead96}
{Readhead}, A.~C.~S., {Taylor}, G.~B., {Pearson}, T.~J., \& {Wilkinson}, P.~N.
  1996, \apj, 460, 634

\bibitem[{{Richards} {et~al.}(2002){Richards}, {Fan}, {Newberg}, {Strauss},
  {Vanden Berk}, {Schneider}, {Yanny}, {Boucher}, {Burles}, {Frieman}, {Gunn},
  {Hall}, {Ivezi{\'c}}, {Kent}, {Loveday}, {Lupton}, {Rockosi}, {Schlegel},
  {Stoughton}, {SubbaRao}, \& {York}}]{Richards02}
{Richards}, G.~T., {Fan}, X., {Newberg}, H.~J., {et~al.} 2002, \aj, 123, 2945

\bibitem[{{Schmidt}(1968)}]{Schmidt68}
{Schmidt}, M. 1968, \apj, 151, 393

\bibitem[{{Schulz} {et~al.}(2015){Schulz}, {Kreikenbohm}, {Kadler}, {Ojha},
  {Ros}, {Stevens}, {Edwards}, {Carpenter}, {Els{\"a}sser}, {Gehrels},
  {Gro{\ss}berger}, {Hase}, {Horiuchi}, {Lovell}, {Mannheim}, {Markowitz},
  {M{\"u}ller}, {Phillips}, {Pl{\"o}tz}, {Quick}, {Tr{\"u}stedt}, {Tzioumis},
  \& {Wilms}}]{Schulz15}
{Schulz}, R., {Kreikenbohm}, A., {Kadler}, M., {et~al.} 2015, ArXiv e-prints
  [\eprint[arXiv]{1511.02631}]

\bibitem[{{Shen} \& {Ho}(2014)}]{Shen14}
{Shen}, Y. \& {Ho}, L.~C. 2014, \nat, 513, 210

\bibitem[{{Son} {et~al.}(2012){Son}, {Woo}, {Kim}, {Fu}, {Kawakatu}, {Bennert},
  {Nagao}, \& {Park}}]{Son12}
{Son}, D., {Woo}, J.-H., {Kim}, S.~C., {et~al.} 2012, \apj, 757, 140

\bibitem[{{Stickel} {et~al.}(1991){Stickel}, {Padovani}, {Urry}, {Fried}, \&
  {Kuehr}}]{Stickel91}
{Stickel}, M., {Padovani}, P., {Urry}, C.~M., {Fried}, J.~W., \& {Kuehr}, H.
  1991, \apj, 374, 431

\bibitem[{{Urry} \& {Padovani}(1991)}]{Urry91}
{Urry}, C.~M. \& {Padovani}, P. 1991, \apj, 371, 60

\bibitem[{{Urry} \& {Padovani}(1995)}]{Urry95}
{Urry}, C.~M. \& {Padovani}, P. 1995, \pasp, 107, 803

\bibitem[{{Urry} \& {Shafer}(1984)}]{Urry84}
{Urry}, C.~M. \& {Shafer}, R.~A. 1984, \apj, 280, 569

\bibitem[{{Wall} \& {Peacock}(1985)}]{Wall85}
{Wall}, J.~V. \& {Peacock}, J.~A. 1985, \mnras, 216, 173

\bibitem[{{Wu}(2009{\natexlab{a}})}]{Wu09b}
{Wu}, Q. 2009{\natexlab{a}}, \apjl, 701, L95

\bibitem[{{Wu}(2009{\natexlab{b}})}]{Wu09}
{Wu}, Q. 2009{\natexlab{b}}, \mnras, 398, 1905

\bibitem[{{Yuan} {et~al.}(2008){Yuan}, {Zhou}, {Komossa}, {Dong}, {Wang}, {Lu},
  \& {Bai}}]{Yuan08}
{Yuan}, W., {Zhou}, H.~Y., {Komossa}, S., {et~al.} 2008, \apj, 685, 801

\bibitem[{{Zhou} {et~al.}(2006){Zhou}, {Wang}, {Yuan}, {Lu}, {Dong}, {Wang}, \&
  {Lu}}]{Zhou06}
{Zhou}, H., {Wang}, T., {Yuan}, W., {et~al.} 2006, \apjs, 166, 128

\end{thebibliography}

\end{document}